\documentclass[%
 reprint,
 superscriptaddress,
 amsmath,amssymb,
 aps,
 pre,
]{revtex4-1}

\usepackage{graphicx}
\usepackage{dcolumn}
\usepackage{bm}
\usepackage[normalem]{ulem}
\usepackage{xcolor}
\usepackage[hidelinks]{hyperref}

\begin{document}


\pagestyle{empty}

\title{Signaling activations through G-protein-coupled-receptor aggregations}

\author{Masaki Watabe}
\thanks{Corresponding author}
\email{masaki@riken.jp}
\affiliation{Laboratory for Biologically Inspired Computing, RIKEN Center for Biosystems Dynamics Research, Suita, Osaka 565-0874, Japan}
\author{Hideaki Yoshimura}
\affiliation{School of Science, The University of Tokyo, 7-3-1 Hongo, Bunkyo-ku, Tokyo 113-0033, Japan}
\author{Satya N. V. Arjunan}
\affiliation{Laboratory for Biologically Inspired Computing, RIKEN Center for Biosystems Dynamics Research, Suita, Osaka 565-0874, Japan}
\affiliation{Lowy Cancer Research Centre, The University of New South Wales, Sydney 2052, Australia}
\author{Kazunari Kaizu}
\affiliation{Laboratory for Biologically Inspired Computing, RIKEN Center for Biosystems Dynamics Research, Suita, Osaka 565-0874, Japan}
\author{Koichi Takahashi}
\thanks{Corresponding author}
\email{ktakahashi@riken.jp}
\affiliation{Laboratory for Biologically Inspired Computing, RIKEN Center for Biosystems Dynamics Research, Suita, Osaka 565-0874, Japan}
\affiliation{Institute for Advanced Biosciences, Keio University, Fujisawa, Kanagawa 252-8520, Japan}


\begin{abstract}
Eukaryotic cells transmit extracellular signal information to cellular interiors through the formation of a ternary complex made up of a ligand (or agonist), G-protein, and G-protein coupled receptor (GPCR). Previously formalized theories of ternary complex formation have mainly assumed that observable states of receptors can only take the form of monomers. Here, we propose a multiary complex model of GPCR signaling activations via the vector representation of various unobserved aggregated receptor states. Our results from model simulations imply that receptor aggregation processes can govern cooperative effects in a regime inaccessible by previous theories. In particular, we show how the affinity of ligand-receptor binding can be largely varied by various oligomer formations in the low concentration range of G-protein stimulus. 
\begin{description}
\item[DOI] \href{https://link.aps.org/doi/10.1103/PhysRevE.102.032413}{\color{blue}10.1103/PhysRevE.102.032413}
\end{description}
\end{abstract}

\maketitle

\section{Introduction}
\begin{figure}[!b]
\centering
\includegraphics[width=0.80\linewidth]{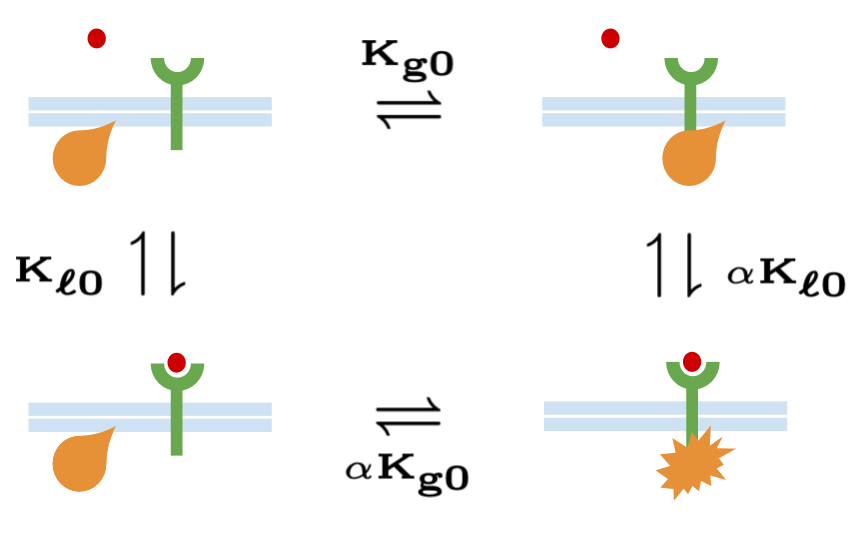}

\caption{Schematic illustration of the simplest ternary complex model~\cite{delean1980, kenakin2017}. A ligand (red bullet) and G-protein (orange object) can bind to a monomeric GPCR (green Y-shaped object) with equilibrium constants $K_{\ell 0}$ and $K_{g0}$, respectively. A ternary complex composed of a ligand, G-protein and receptor can be formed in two ways: ($i$) ligands can interact with the receptors binding to G-proteins with equilibrium constant $\alpha K_{\ell0}$, and ($ii$) G-proteins can bind to the ligand-bound receptors with equilibrium constant $\alpha K_{g0}$, where $\alpha$ is a cooperativity factor that denotes the mutual effect of the receptor-binding affinity to the ligand and G-protein.}
\label{fig01;tcm}
\end{figure}

G-protein coupled receptors (GPCR) in eukaryotic cells form a remarkable modular system over the cell membrane, its main function being to provide cells with a wide means of signal communications between extracellular molecules (e.g., hormones and neurotransmitters) and intracellular signaling G-proteins. GPCR signal communication can be achieved through conformational changes in receptors, as well as the complex formation with three different components: a ligand (or agonist), G-protein, and GPCR. This complex formation serves as an activated signaling component, accordingly changing the affinity of ligand-receptor binding as a function of G-protein stimulus. These ideas have been formalized into mechanistic theories of GPCR signaling activation, termed ternary complex models~\cite{delean1980, kenakin2017}, following in particular, the assumption that receptors can only take the form of monomers (see Figure~\ref{fig01;tcm}).

%

Thousands of GPCRs diffuse on the cell membrane, randomly interacting with each other and spontaneously forming oligomers such as dimers and trimers. The dimer formations of ligand-bound receptors may, for example, extend the colocalization period of the activated signaling component~\cite{nishiguchi2020}. Likewise, in a wide concentration range of ligand stimulus, group behavior such as cooperativity induced by receptor dimerization may be constrained by the affinity of higher-order oligomer formations~\cite{watabe2019, hiroshima2018, hiroshima2012, *hiroshima2013, teramura2006, uyemura2005, wofsy1992a, *wofsy1992b}. Recent experimental studies of receptor systems have shown the existence and functionality of dimerization and higher-order oligomerization of receptors, implying modifications to ternary complex models~\cite{nishiguchi2020, watabe2019, hiroshima2018, hiroshima2012, *hiroshima2013, teramura2006, uyemura2005, wofsy1992a, *wofsy1992b, yanagawa2018, *yanagawa2011a, *yanagawa2011b, guo2017, milligan2007, breitwieser2004, bai2004}. This lies in contrast to previous theories which have mostly focused on scenarios where receptor-receptor couplings (i.e., direct interaction of two receptors) are weakly linked with GPCR signaling activations.


\begin{figure}
\leftline{\bf (a)}
\centering
\includegraphics[width=0.80\linewidth]{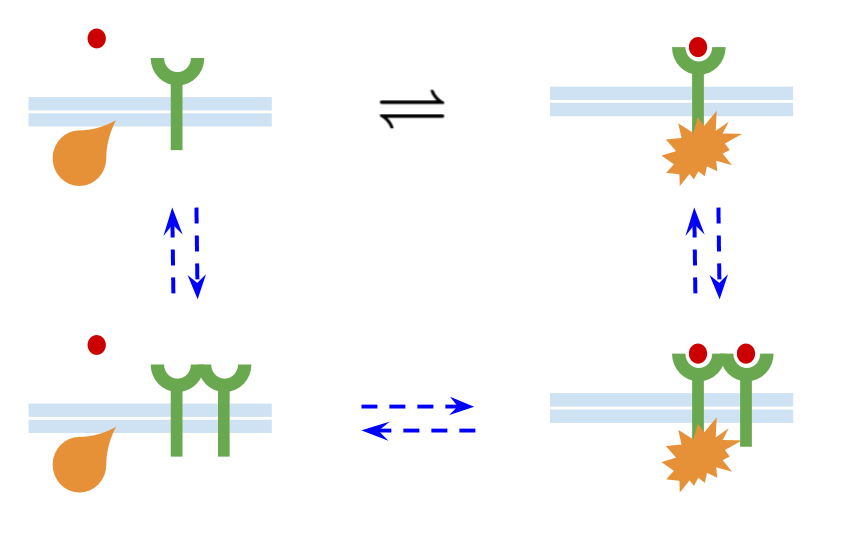}

\leftline{\bf (b1)}
\centering
\includegraphics[width=0.50\linewidth]{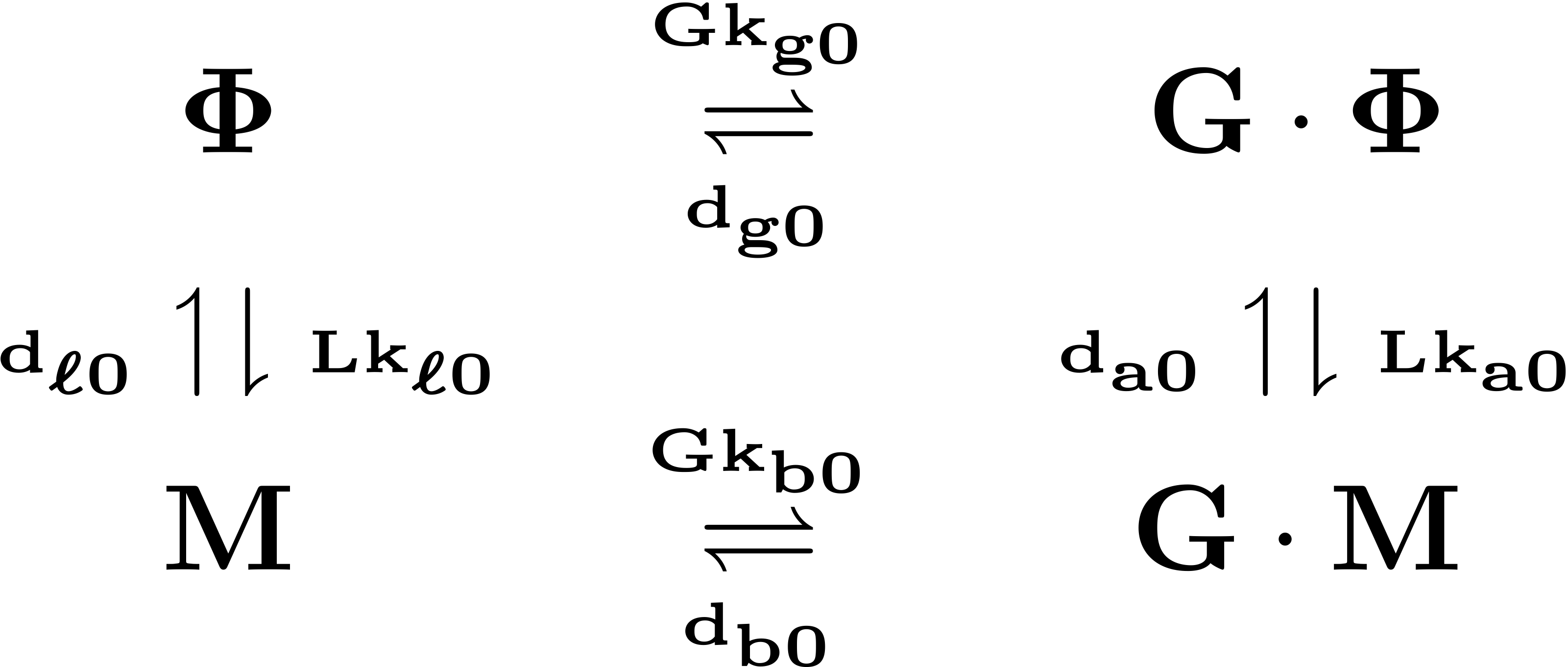}

\leftline{\bf (b2)}
\centering
\includegraphics[width=0.80\linewidth]{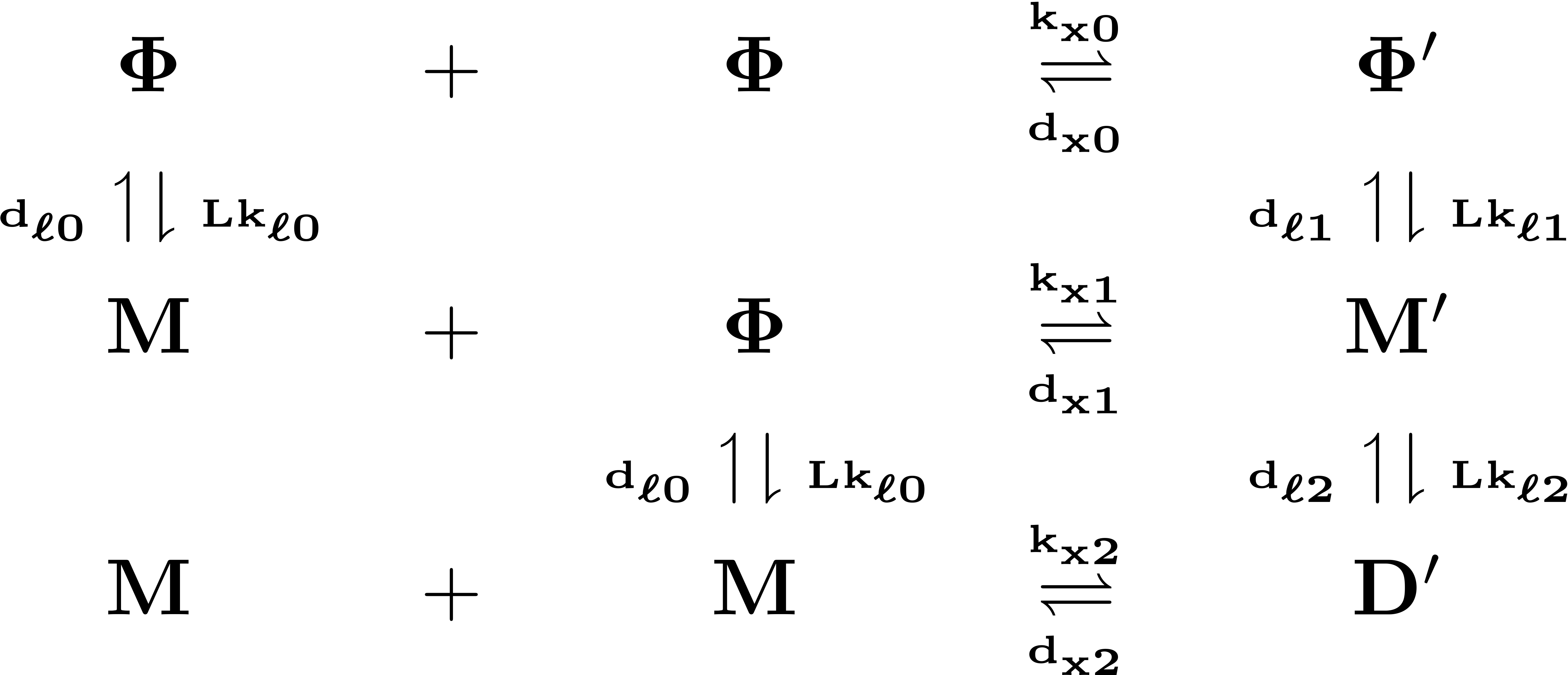}

\leftline{\bf (b3)}
\centering
\includegraphics[width=0.54\linewidth]{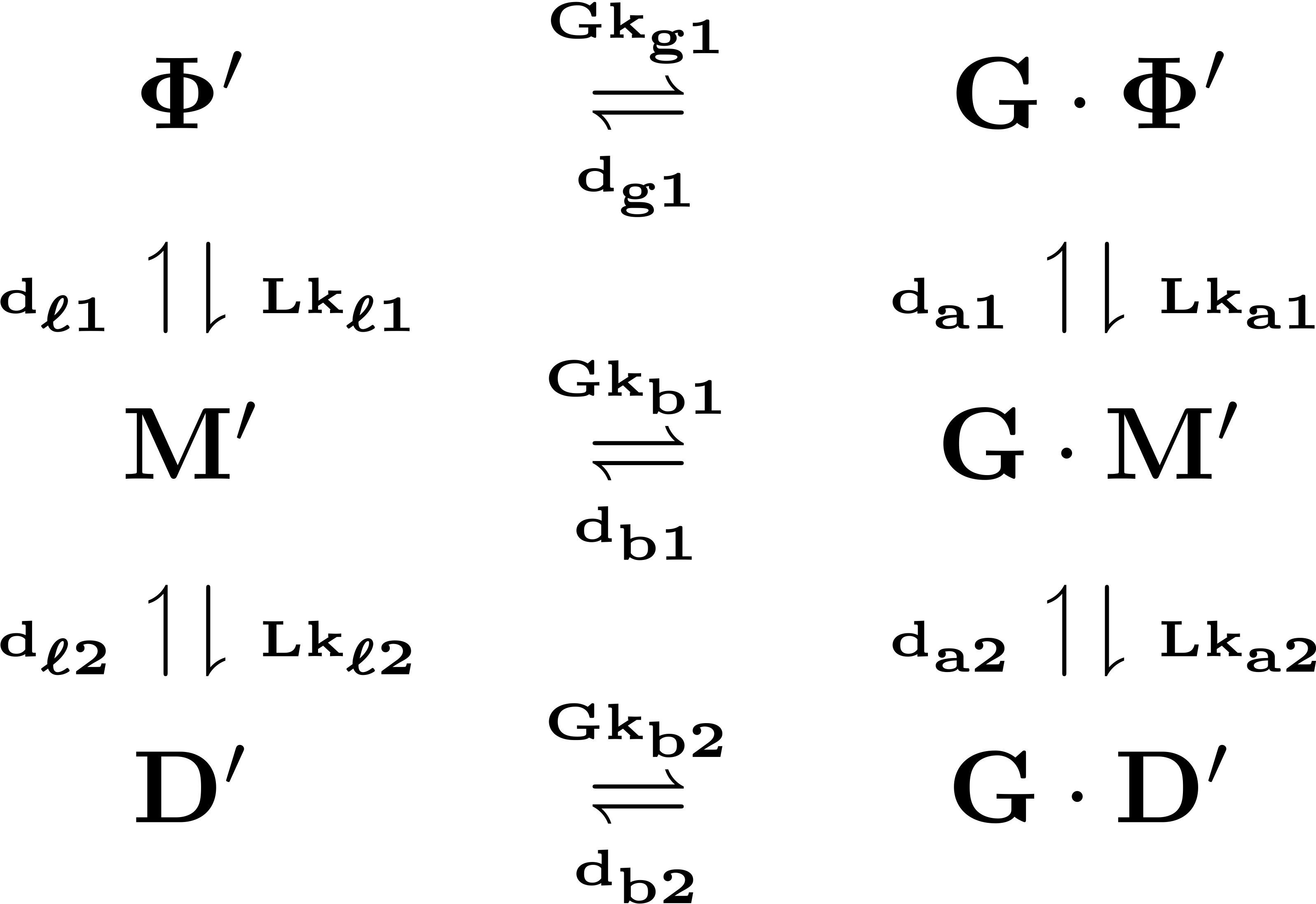}

\leftline{\bf (b4)}
\centering
\includegraphics[width=0.80\linewidth]{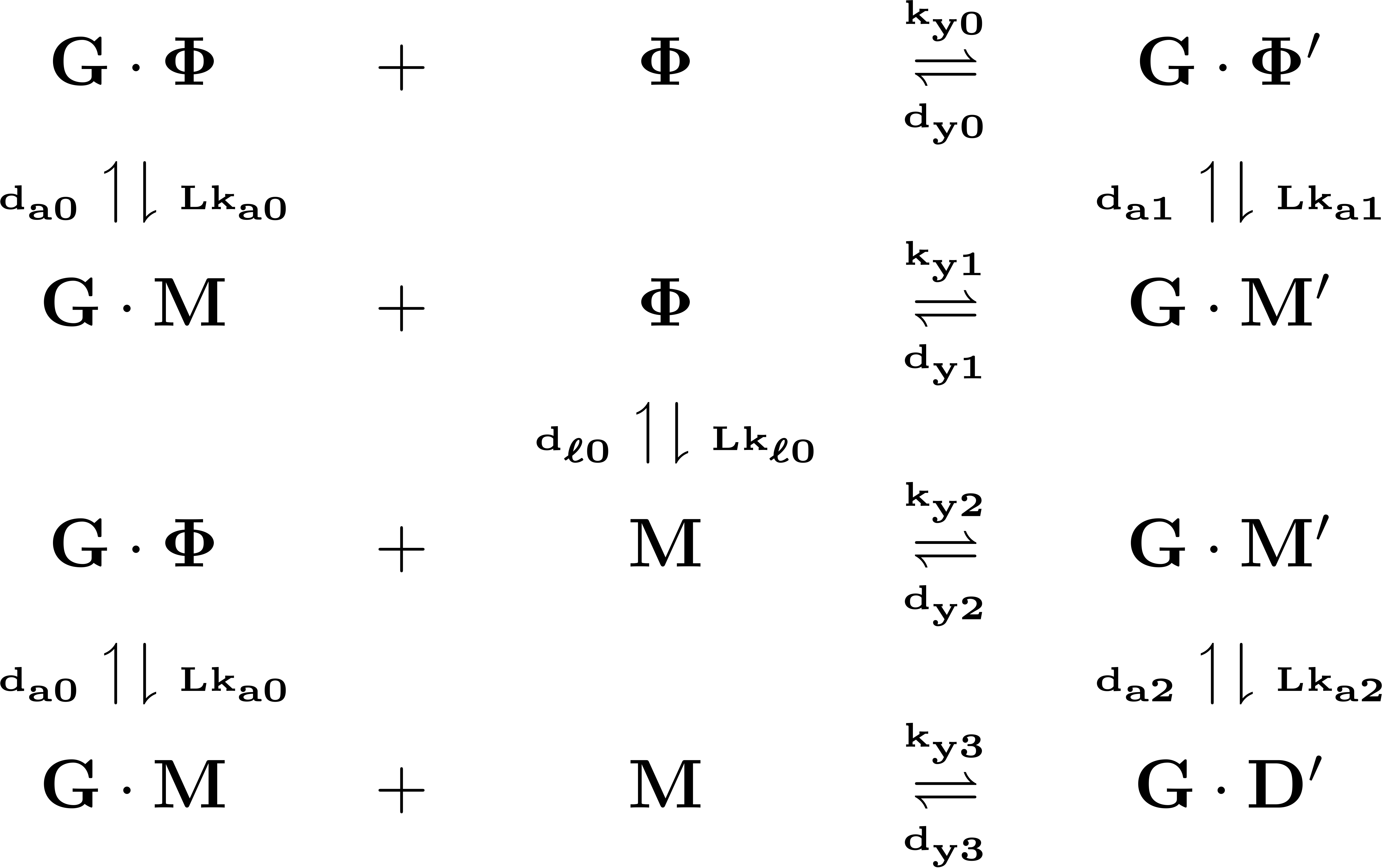}

\caption{Multiary complex formation via ligand-bound receptor dimerization in GPCR signaling activations. (a) Schematic illustration of the ternary complex formation (black arrows) and receptor aggregation processes (blue arrows)~\cite{nishiguchi2020}. The multiary complex is composed of a ligand (red bullet), G-protein (orange object), and various aggregated receptors (green Y-shaped objects). (b) Network diagrams of the multiary complex model. (b1) Multivalent form of the simplest ternary complex model. (b2) Dimer formations of ligand-bound receptors. (b3) First-order interactions of ligand and G-protein to receptor states. (b4) Second-order interactions of receptor states to G-protein-bound receptor states. $\bf\color{blue}L$ and $\bf\color{red}G$ represent the input concentrations of ligand and G-protein stimuli, respectively. See the main text for detailed descriptions of state vectors and model parameters.}
\label{fig02;network}
\end{figure}


A key challenge to model modification is the incorporation of realistic but unobserved receptor aggregations into model network. Ternary complex models have been constructed with various biochemical parameters such as ligand-receptor binding rates (see Figure~\ref{fig01;tcm}) but restricted to the observable receptor states imposed by experimental techniques (e.g., live-cell imaging via biomolecules tagged with fluorescent emitters). While such models mainly assume that observable states of receptors can only take the form of monomers, effects arising from oligomer formations of unobserved receptors have received less attention. In this article, we propose a multiary (or $n$-ary) complex model of GPCR signaling activations via dimer formations of ligand-bound receptors, represented by a multivalent form of physical observables under basis vectors of various unobserved aggregated receptor states (see Figure~\ref{fig02;network}). We then perform model simulations to explore the biophysical effects of receptor aggregation in GPCR signaling activations. Crucially, we show how a mixture of various unobserved aggregated receptor states can lead to the transition of ligand-receptor binding affinity in a regime which cannot be predicted by ternary complex models. We finally promote a further modification to the multiary complex models, including in particular, interactions between inactive and active states of the receptor observables. Such model modification is of broad relevance beyond just receptor aggregation presented here, possibly leading to a more general modeling framework of GPCR signaling activations.

\section{Multiary complex model}
\subsection{Model framework}
The receptor state vector representation of physical observables provides a concrete organizational framework to link model components (e.g., ligand, receptor and G-protein) with physical observables imposed by experimental techniques (e.g., single molecule imaging via ligands fused with tetramethylrhodamine). In this vector representation, the multiary complex model is described by a function containing the probabilities of biochemical interactions that form various unobserved aggregated receptor states. All possible aggregated receptor states via dimer formation of ligand-bound receptors can be treated mathematically as basis vectors in a multidimensional real vector space.
 
First, we assume that receptor states are physically observable if a ligand binds to a receptor. In the case that no ligand binds to a receptor, receptor states are physically unobservable or null. Figure~\ref{fig02;network}b shows network diagrams of the multiary complex model. Null (${\bf \Phi}$ and ${\bf \Phi'}$), monomeric (${\bf M}$ and ${\bf M'}$) and dimeric (${\bf D'}$) observable state vectors of receptors are given by
\begin{equation}
{\bf \Phi} = \left(\begin{array}{c}r \\rr \\rrr \\\vdots \\r^{^{N}} \end{array}\right),
\ \ 
{\bf M} = \left(\begin{array}{c}R \\Rr \\Rrr \\\vdots \\Rr^{^{N-1}} \end{array}\right),
\ \ 
{\bf \Phi'} = \left(\begin{array}{c}r\cdot r \\r\cdot rr \\r\cdot rrr \\\vdots \\r^{^{N}}\cdot r^{^{N}} \end{array}\right),
\nonumber
\end{equation}
\begin{equation}
{\bf M'} = \left(\begin{array}{c}R\cdot r \\R\cdot rr \\R\cdot rrr \\\vdots \\Rr^{^{N-1}}\cdot r^{^{N}} \end{array}\right),
\ \ 
{\bf D'} = \left(\begin{array}{c}R\cdot R \\R\cdot Rr \\R\cdot Rrr \\\vdots \\Rr^{^{N-1}}\cdot Rr^{^{N-1}} \end{array}\right)
\end{equation}
where $r$ and $R$ represent the receptors and the ligand-bound receptors, respectively; $r \cdot r = rr$ denotes the dimer of two free receptors. $N$ refers to the number of receptors that can be aggregated in the ${\bf \Phi}$ and ${\bf M}$ observable states. There are $N^2$ elements in the ${\bf \Phi'}$, ${\bf M'}$, and ${\bf D'}$ observable states. 

Null (${\bf G\Phi}$ and ${\bf G\Phi'}$), monomeric (${\bf GM}$ and ${\bf GM'}$) and dimeric (${\bf GD'}$) observable state vectors of the G-protein-bound receptors are also given by
\begin{equation}
{\bf G\Phi} = \left(\begin{array}{c}G \cdot r \\G \cdot rr \\G \cdot rrr \\\vdots \\G \cdot r^{^{N}} \end{array}\right),
\ \ 
{\bf GM} = \left(\begin{array}{c}G \cdot R \\G \cdot Rr \\G \cdot Rrr \\\vdots \\G \cdot Rr^{^{N-1}} \end{array}\right),
\nonumber
\end{equation}
\begin{equation}
{\bf G\Phi'} = \left(\begin{array}{c}G \cdot r\cdot r \\G \cdot r\cdot rr \\G \cdot r\cdot rrr \\\vdots \\G \cdot r^{^{N}}\cdot r^{^{N}} \end{array}\right),
\ \ 
{\bf GM'} = \left(\begin{array}{c}G \cdot R\cdot r \\G \cdot R\cdot rr \\G \cdot R\cdot rrr \\\vdots \\G \cdot Rr^{^{N-1}}\cdot r^{^{N}} \end{array}\right),
\nonumber
\end{equation}
\begin{equation}
{\bf GD'} = \left(\begin{array}{c}G \cdot R\cdot R \\G \cdot R\cdot Rr \\G \cdot R\cdot Rrr \\\vdots \\G \cdot Rr^{^{N-1}}\cdot Rr^{^{N-1}} \end{array}\right)
\end{equation}
where $G \cdot r$ and $G \cdot R$ represent G-protein-bound receptors. $N$ refers to the number of receptors that can be aggregated in the ${\bf G\Phi}$, and ${\bf GM}$ observable states. There are $N^2$ elements in the ${\bf G\Phi'}$, ${\bf GM'}$, and ${\bf GD'}$ observable states. 

The receptor state vector representation allows for the arrangement of the association and dissociation rates of higher-order oligomers into matrix representations. In first-order interactions (e.g., $\bf\Phi \rightleftharpoons M$) of a ligand and G-protein to a receptor, and the rates of association (${\bf k_{\it p}}$) and dissociation (${\bf d_{\it p}}$) of the $p$-th index are represented by $N \times N$ (or $N^2 \times N^2$) diagonal matrices acting upon the basis vectors, transforming an aggregated state into an observable state. These diagonal matrices can be written in the form of
\begin{equation}
{\bf k_{\it p}} = \left(
\begin{array}{ccccc}
k_{p,0} & 0 & \cdots & 0 \\
0 & k_{p,1} & \cdots & 0 \\
\vdots & \vdots & \ddots & \vdots \\
0 & 0 & \cdots & k_{p,N (or N^2)}
\end{array}
\right),
\end{equation}
\begin{equation}
{\bf d_{\it p}} = \left(
\begin{array}{ccccc}
d_{p,0} & 0 & \cdots & 0 \\
0 & d_{p,1} & \cdots & 0 \\
\vdots & \vdots & \ddots & \vdots \\
0 & 0 & \cdots & d_{p,N (or N^2)}
\end{array}
\right)
\end{equation}
where $p = \ell0$, $\ell1$, $\ell2$, $g0$, $g1$, $g2$, $a0$, $a1$, $a2$, $b0$, $b1$ and $b2$. 

The dissociation rates (${\bf d_{\it q}}$) of the $q$-th index for second-order interactions (e.g., $\bf\Phi + M \rightleftharpoons M'$) in Figures~\ref{fig02;network}b1 and b4 are represented by $N^2 \times N^2$ diagonal matrices. Non-diagonal matrices of the association rates (${\bf k_{\it q}}$) of the $q$-th index can, however, transform a mixture of various aggregated states into an observable state. These matrices are given by
\begin{equation}
{\bf k_{\it q}} = \left(
\begin{array}{ccccc}
k_{q,00} & k_{q,01} & \cdots & k_{q,0N} \\
k_{q,10} & k_{q,11} & \cdots & k_{q,1N} \\
\vdots & \vdots & \ddots & \vdots \\
k_{q,N0} & k_{q,N1} & \cdots & k_{q,NN}
\end{array}
\right),
\end{equation}
\begin{equation}
{\bf d_{\it q}} = \left(
\begin{array}{ccccc}
d_{q,0} & 0 & \cdots & 0 \\
0 & d_{q,1} & \cdots & 0 \\
\vdots & \vdots & \ddots & \vdots \\
0 & 0 & \cdots & d_{q,N^2}
\end{array}
\right)
\end{equation}
where $q = x0$, $x1$, $x2$, $y0$, $y1$, $y2$ and $y3$.

\subsection{Multivalent cell models}
First, we construct multivalent ($N = 1, 2, 3, 4$, and $5$) cell models of multiary complex formations. We then use the E-cell system version 4~\cite{kaizu2019, *chew2019, *chew2018} to simulate cell models of biological fluctuation that arise from stochastic changes in the cell surface geometry, number of receptors, ligand binding, molecular states, and diffusion constants. These cell models assume that non-diffusive receptors are uniformly distributed on the cell membrane. The source code of the multivalent cell models is provided in the Supplemental Material~\cite{sm}. 


In particular, we assume that the diagonal matrices in the first-order association (${\bf k_{\it p}}$) and dissociation rates (${\bf d_{\it p}}$) of the $p$-th index are given by
\begin{equation}
{\bf k_{\it p}} = k_{p} \boldsymbol{\mathcal{I}},
\ \ \ 
{\bf d_{\it p}} = d_{p} \boldsymbol{\mathcal{I}}
\end{equation}
where $p = \ell0$, $\ell1$, $\ell2$, $g0$, $g1$, $g2$, $a0$, $a1$, $a2$, $b0$, $b1$ and $b2$. $\boldsymbol{\mathcal{I}}$ represents an $N \times N$ (or $N^2 \times N^2$) identity matrix where every diagonal element is equal to one, but every off-diagonal element is zero. Also, the second-order association ($\bf k_{\it q}$) and dissociation rates ($\bf d_{\it q}$) of the $q$-th index can be written in the matrix form of  
\begin{equation}
{\bf k_{\it q}} = \frac{k_{q}}{N} \boldsymbol{\mathcal{J}},
\ \ \ 
{\bf d_{\it q}} = d_{q} \boldsymbol{\mathcal{I}}
\end{equation}
where $q = x0$, $x1$, $x2$, $y0$, $y1$, $y2$ and $y3$. $\boldsymbol{\mathcal{J}}$ represents an $N \times N$ all-ones matrix where every element is equal to one. $\boldsymbol{\mathcal{I}}$ also denotes an $N^2 \times N^2$ identity matrix.

\begin{figure*}
\leftline{\bf \hspace{0.06\linewidth} (a) \hspace{0.28\linewidth} (b) \hspace{0.25\linewidth} (c)}
\centering
     \includegraphics[width=0.29\linewidth]{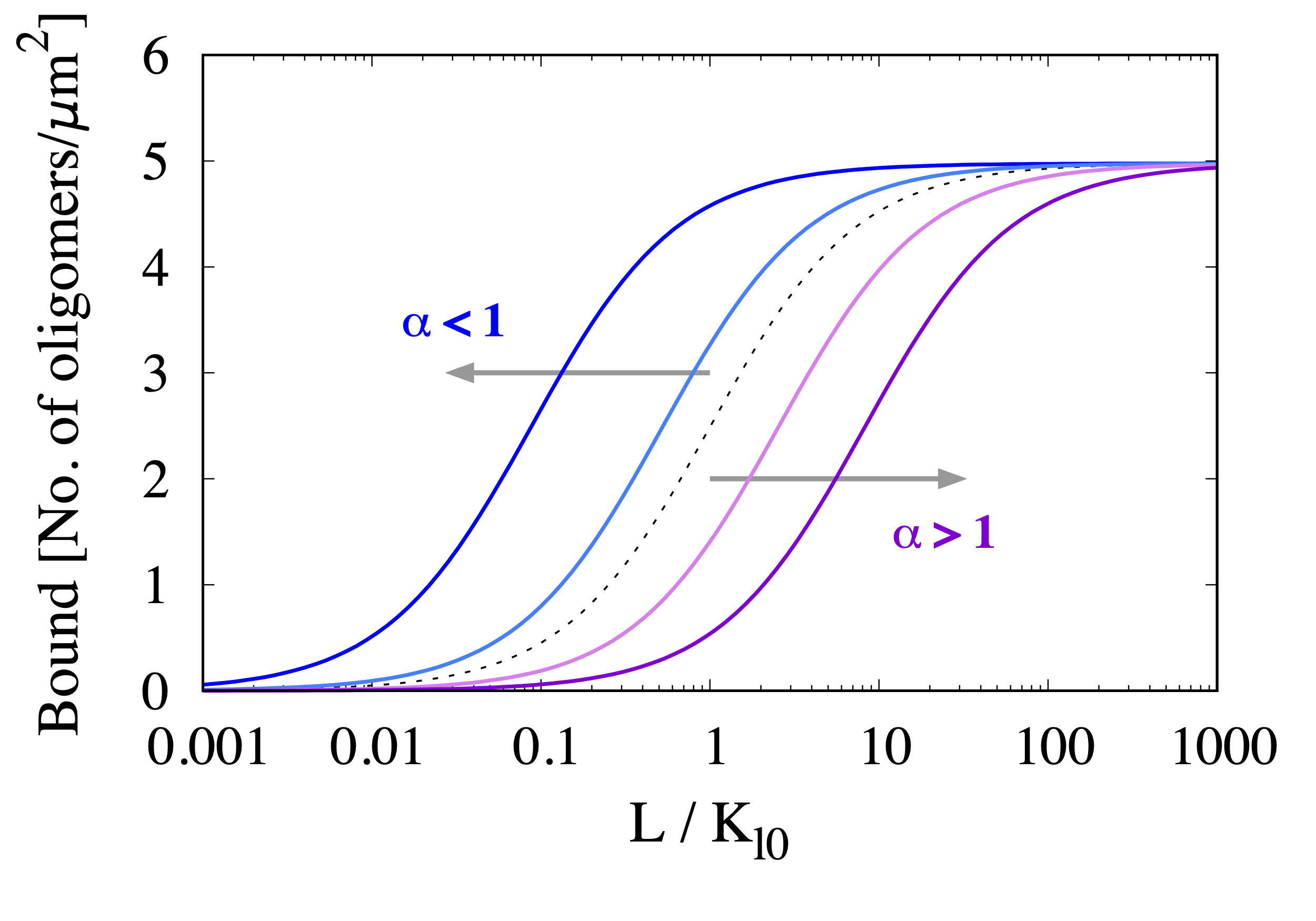}
     \includegraphics[width=0.30\linewidth]{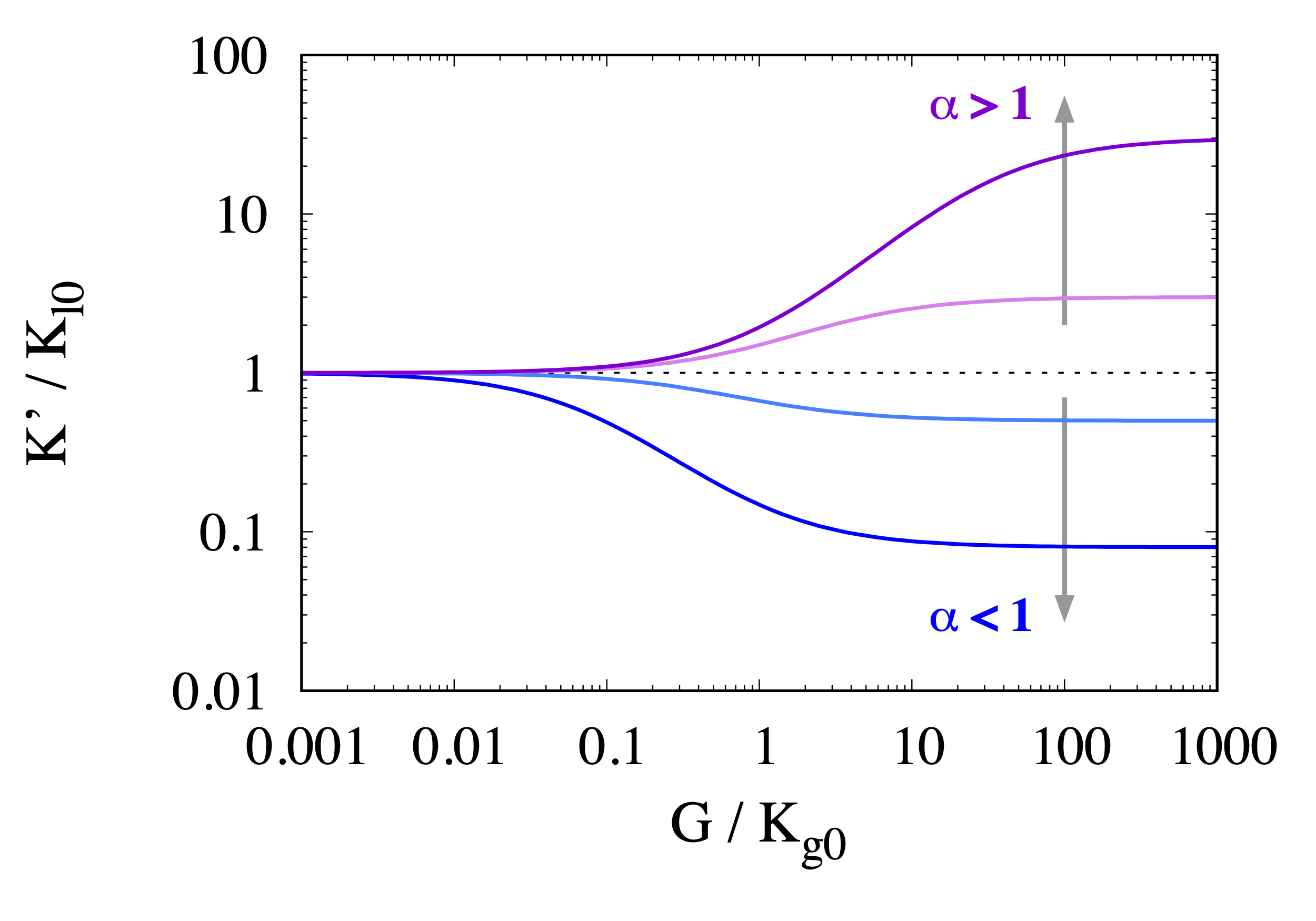}
     \includegraphics[width=0.30\linewidth]{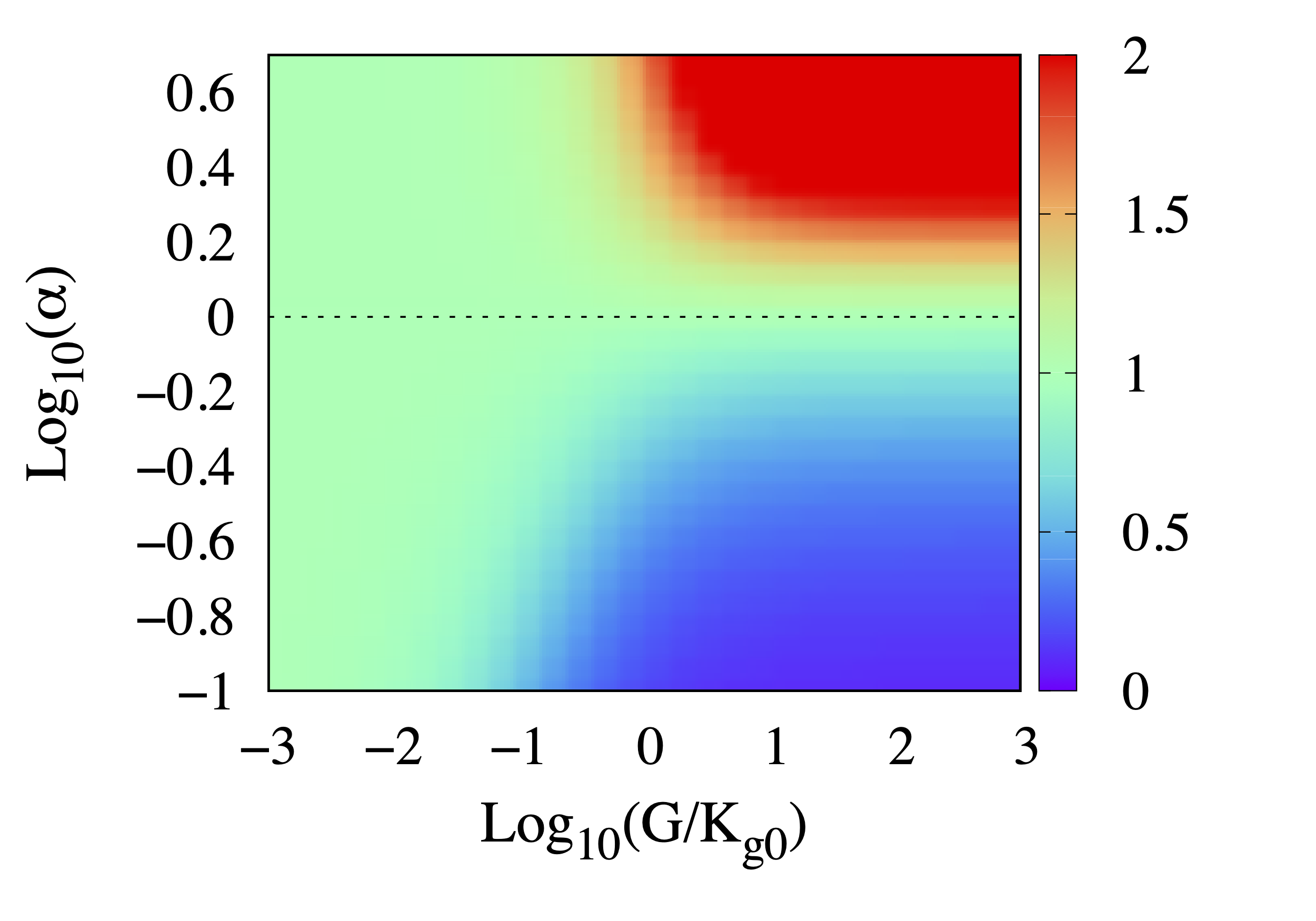}

\leftline{\bf \hspace{0.06\linewidth} (d) \hspace{0.28\linewidth} (e) \hspace{0.25\linewidth} (f)}
\centering
     \includegraphics[width=0.29\linewidth]{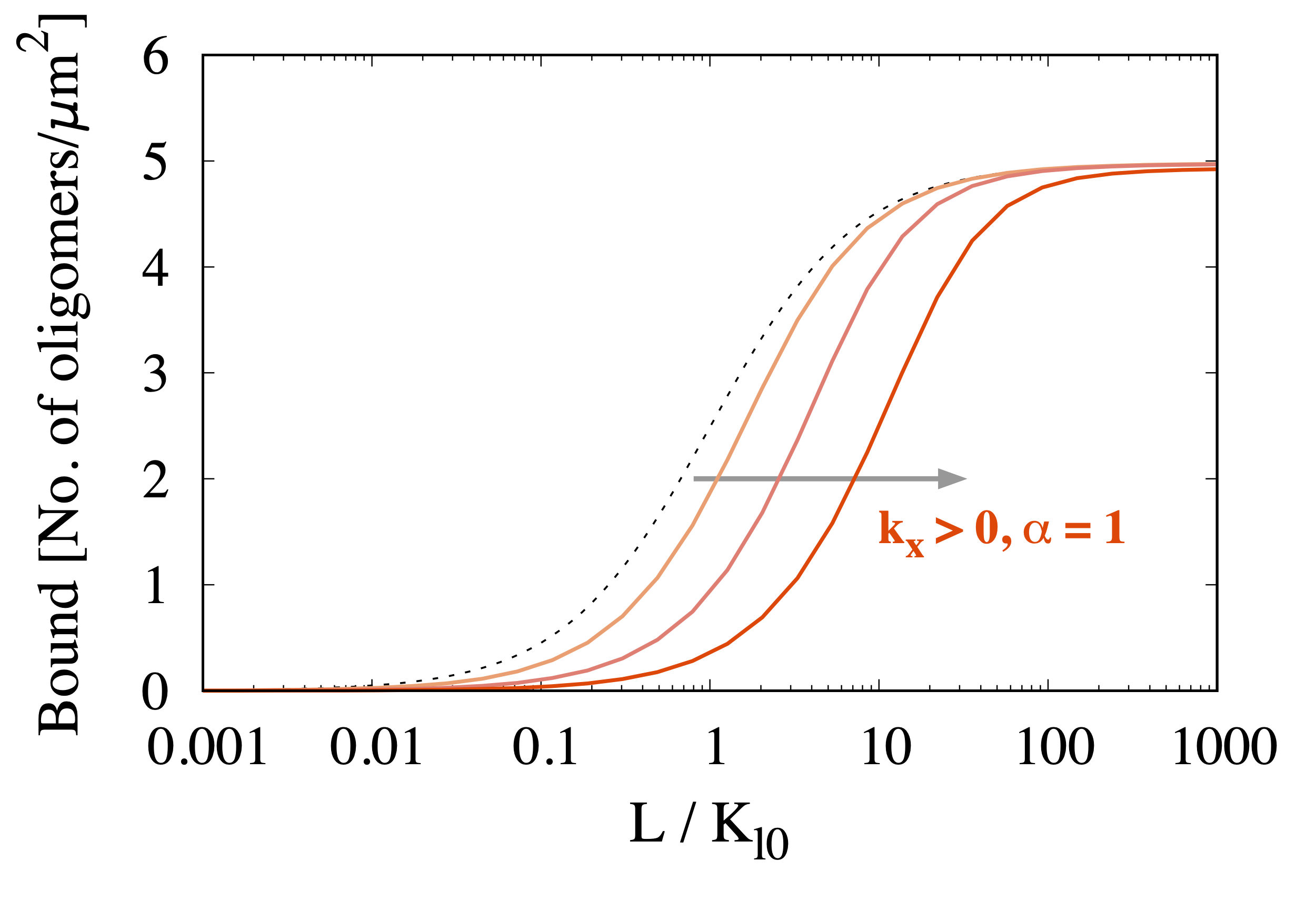}
     \includegraphics[width=0.30\linewidth]{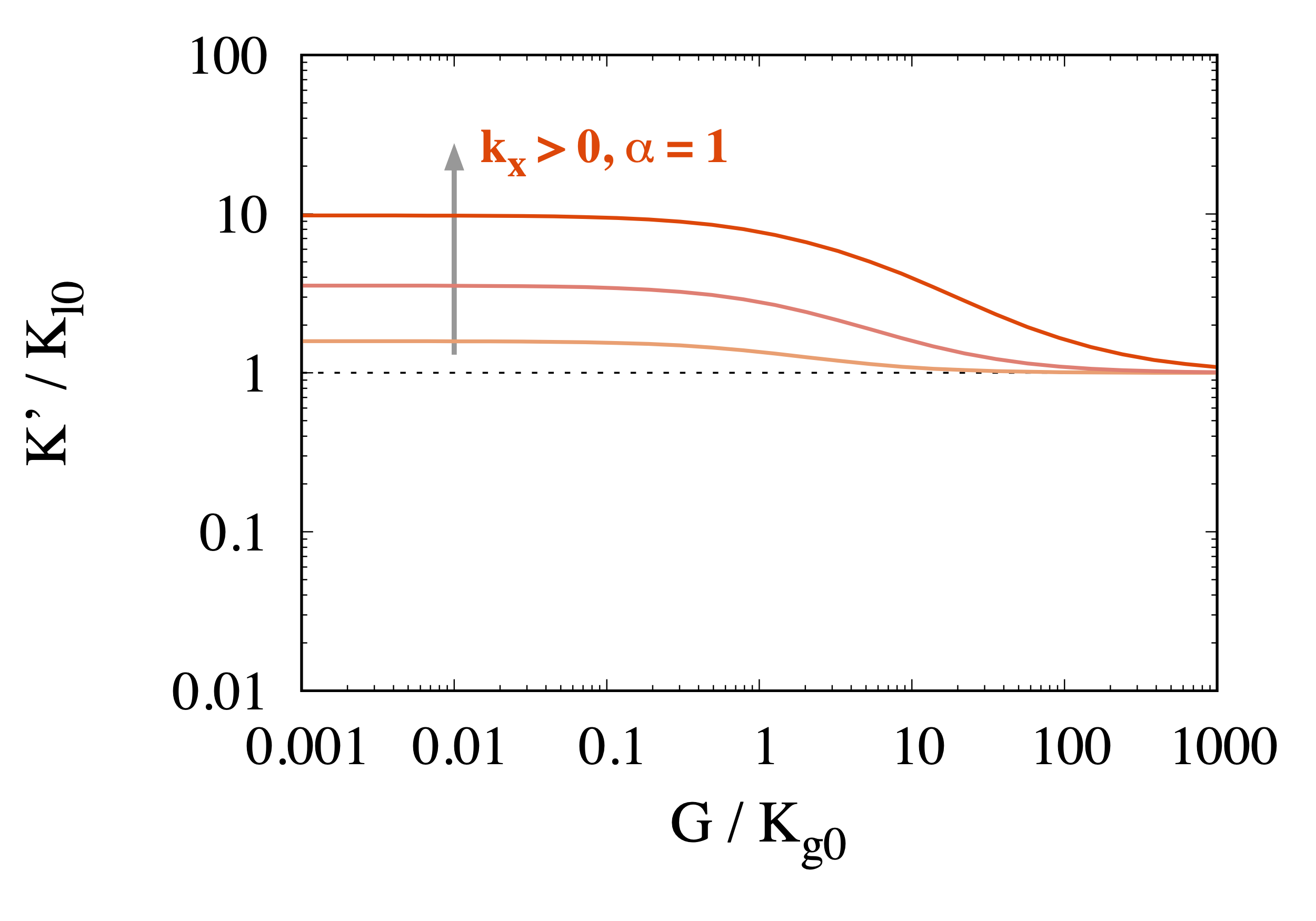}
     \includegraphics[width=0.30\linewidth]{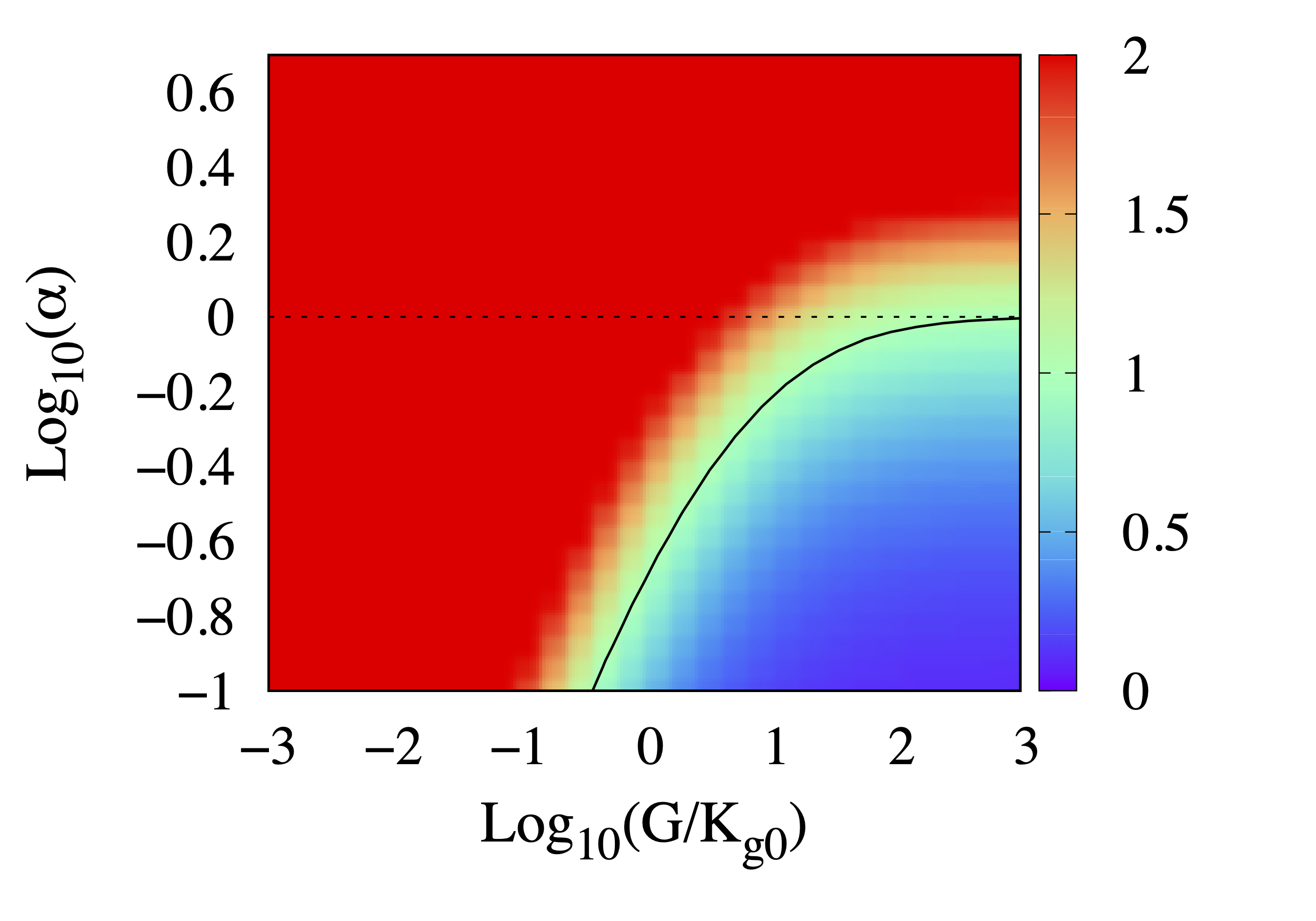}

\caption{Model comparison between the simplest ternary complex model (top row) and the monovalent cell model (bottom row). $L$ and $G$ represent the input concentrations of ligand and G-protein stimuli, respectively. (a) Ligand-receptor binding curve as a function of ligand stimulus $L$, is given by Eq.~(\ref{eqn;binding_tcm}) where $G/K_{g0} = 0.1$, $\alpha = 0.08$ (blue), $0.5$ (light-blue), $1.0$ (black dots), $3.0$ (light-violet), and $30$ (violet). (b) Overall affinity of ligand-receptor binding ($K'/K_{\ell 0}$) as a function of G-protein stimulus, is given by the Eq.~(\ref{eqn;affinity_tcm}), for $G/K_{g0} = 0.1$ and $\alpha = 0.08$, $0.5$, $1.0$, $3.0$, $30$. (c) Colors represent the binding affinity of the ternary complex model in the range from $0$ (blue) to $\sim1$ (green) and $\ge2$ (red); the black dashed line shows $K'/K_{\ell 0} = 1$. (d) For $\alpha = 1$ and $k_x = 0$ (black dots), $10^{-4}$ (light-orange), $10^{-3}$ (orange), and $10^{-2}$ (dark-orange), ligand-receptor binding curves are shown as a function of ligand stimulus $L/K_{\ell0}$. (e) For $\alpha = 1$ and $k_x = 0$, $10^{-4}$, $10^{-3}$, and $10^{-2}$, overall affinities of ligand-receptor binding are shown as a function of G-protein stimulus $G/K_{g0}$. (f) Color represents the binding affinity of the monovalent cell model as a function of $\alpha$ and $G/K_{g0}$, assuming $k_x = 0.001$; the black dashed and solid lines represent $K'/K_{\ell 0} = 1$ in the simplest ternary complex model and the monovalent cell model, respectively.}
\label{fig03;comparison} 
\end{figure*}

The model parameter values are given as follows: total receptor concentration, $T = 4.977\ {\rm receptors/\mu m^{2}}$; cooperativity factor, $\alpha = K_{a0}/K_{\ell 0} = K_{b0}/K_{g0}$; binding affinity and dissociation rates for each first-order interactions, $K_{\ell 1} = K_{\ell 2} = 100 K_{\ell 0}$, $K_{g1} = 100 K_{g0}$, $K_{a1} = K_{a2} = 100 K_{a0}$, $K_{b1} = K_{b2} = 100 K_{b0}$, $d_{\ell 0} = d_{\ell 1} = d_{\ell 2} = d_{g0} = d_{g1} = d_{a0} = d_{a1} = d_{a2} = d_{b0} = d_{b1} = d_{b2} = d_{b3} = 1.00\ {\rm s^{-1}}$; and binding affinity and dissociation rates for each second-order interactions, $K_{y0} = K_{x0}$, $K_{x1} = K_{y1} = K_{y2} = K_{\ell 1} K_{x0}/K_{\ell 0}$, $K_{x2} = K_{y3} = K_{\ell 2} K_{\ell 1} K_{x0}/K^2_{\ell 0}$, $d_{x0} = d_{x1} = d_{x2} = d_{y0} = d_{y1} = d_{y2} = d_{y3} = 1.00\ {\rm s^{-1}}$. The local equilibrium constants for the association and dissociation rates satisfy the relation $K_{m} = d_{m}/k_{m}$ where $m = \ell0$, $\ell1$, $\ell2$, $g0$, $g1$, $g2$, $a0$, $a1$, $a2$, $b0$, $b1$, $b2$, $x0$, $x1$, $x2$, $y0$, $y1$, $y2$ and $y3$.

For convenience, we define a dimensionless lumped parameter that depends on the second-order interaction rates of null observables ($\bf\Phi + \Phi \rightleftharpoons  \Phi'$). The lumped parameter can be written in the matrix form of
\begin{equation}
{\bf k_{x}} = \frac{k_{x}}{N} \boldsymbol{\mathcal{J}}
\end{equation}
where $k_x = T/K_{x0}$.

In a concentration range of ligand stimulus ($L/K_{\ell 0}$) from $10^{-3}$ to $10^{3}$, we ran model simulations for a period of $100,000\ {\rm s}$ to verify the complete convergence of receptor response to full equilibrium. To characterize the overall ligand-receptor binding affinities of the multivalent cell models, the Hill function can be fitted to the ligand-receptor binding curves of the monomeric and dimeric observable state vectors of receptors: $\bf M$, $\bf GM$, $\bf M'$, $\bf GM'$, $\bf D'$ and $\bf GD'$. The Hill function can generally be written in the form of
\begin{eqnarray}
B(L) & = & \frac{B_0 L^n}{L^n + K'^n}
\label{eqn;binding_tcm}
\end{eqnarray}
where $L$, $B_0$, and $n$ represent ligand concentration, maximum area-density of the ligand-bound receptor, and the Hill-coefficient, respectively. $K'$ denotes  overall affinity of ligand-receptor binding as a function of G-protein stimulus ($G/K_{g0}$) from $10^{-3}$ to $10^{3}$.

\subsection{The simplest ternary complex model}
The network diagram of the multiary complex model  (see Figure~\ref{fig02;network}b) converges to that of the simplest ternary complex model (see Figure~\ref{fig01;tcm}) as $N = 1$ and $k_{x} \to 0$. Overall ligand-receptor binding states ($\bf M$ and $\bf GM$) in the ternary complex model can be written in the form of
\begin{eqnarray}
B(L) & = & \frac{B_0 L}{L + K'}
\label{eqn;binding_tcm}
\end{eqnarray}
where $L$ and $B_0$ represent ligand concentration and maximum area-density of the ligand-bound receptor, respectively. Overall affinity of ligand-receptor binding as a function of G-protein stimulus is given by
\begin{equation}
K' = K_{\ell 0} \left( \frac{1 + G/K_{g0}}{1 + G/\left(\alpha K_{g0}\right)} \right)
\label{eqn;affinity_tcm}
\end{equation}
where $G$ and $\alpha$ represent G-protein concentration and the cooperativity factor that satisfies the relation $\alpha = K_{a0}/K_{\ell 0} = K_{b0}/K_{g0}$, respectively~\cite{kenakin2017}. If $\alpha = 1$, there is no affinity transition.

\section{Results}
We compare the ligand-receptor binding curves between the simplest ternary complex model ($k_x = 0$) and the monovalent cell model ($k_x > 0$ and $N = 1$). Figure~\ref{fig03;comparison} clearly shows differences in the binding curves. In the simplest ternary complex model, the cooperativity factor $\alpha$ in Eq.~(\ref{eqn;affinity_tcm}) plays a key role in largely varying the overall affinities of ligand-receptor binding in the high concentration range of G-protein stimulus (see Figures~\ref{fig03;comparison}a,b, and c). The affinities of the binding curves shown in Figures~\ref{fig03;comparison}a and b, can be decreased ($K' < K_{\ell0}$) or increased ($K' > K_{\ell0}$) as a function of the cooperativity factor ($\alpha$). Such affinity transitions are also shown in Figure~\ref{fig03;comparison}c, varying from blue to red regions in the high $G/K_{g0}$ range.

To see the effects arising from the second-order interactions in the monovalent cell model, the lumped parameter ($k_x$) can be varied from $0$ to $10^{-2}$, assuming $\alpha = 1$. Figures~\ref{fig03;comparison}d and e, show that the overall affinity of the ligand-receptor binding curves in the low concentration range of G-protein stimulus can be increased ($K' > K_{\ell0}$) as a function of $k_x$. Such affinity shifts are also shown in Figure~\ref{fig03;comparison}f, represented by red colored region in the low $G/K_{g0}$ range. In the absence of G-protein stimulus ($G \to 0$), the overall network diagram of the multiary complex model converges to the dimer formations of ligand-bound receptors (see Figure~\ref{fig02;network}b). In particular, the dimerization model satisfies a specific parameter condition that gives rise to positive cooperativity ($K_{\ell1} = K_{\ell2}$), increasing the overall affinity of ligand-receptor binding~\cite{watabe2019, wofsy1992a}. Because of this cooperative characteristic, the monovalent cell model displays the affinity transition in the low $G/K_{g0}$ range.

\begin{figure}
\centering
     \includegraphics[width=0.80\linewidth]{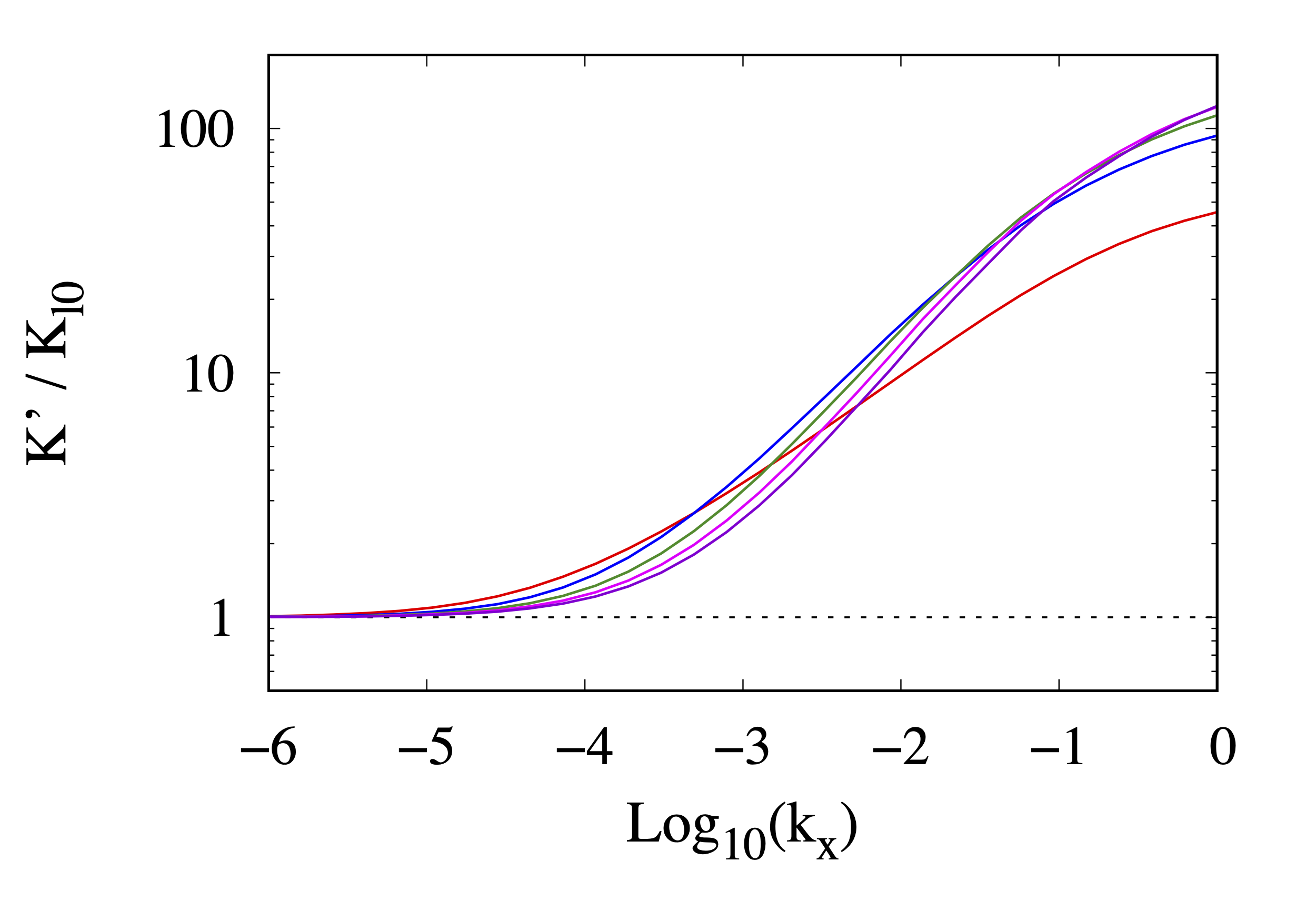}

\caption{Transition of the overall affinity in the multivalent cell models is shown as a function of the lumped parameter $k_x$, assuming $\alpha = 1$ and $G/K_{g0} = 10^{-3}$. Each colored line represents the monovalent (red), bivalent (blue), trivalent (green), tetravalent (pink), and pentavalent (violet) models; the dashed black line denotes $K'/K_{\ell0} = 1$.}
\label{fig04;affinity}
\end{figure}

The affinity transitions can be also seen in higher-order multivalent cell models: bivalent ($N=2$), trivalent ($N=3$), tetravalent ($N=4$), and pentavalent ($N=5$). For $\alpha = 1$ and $G/K_{g0} = 10^{-3}$, Figure~\ref{fig04;affinity} shows the affinity transitions in multivalent cell models as a function of the lumped parameter $k_x$. While the affinity in the multivalent cell models is always unity if $k_x = 0$ ($K' = K_{\ell0}$; black dashed line), the affinity can be increased through the increase of  $k_x$ ($K' > K_{\ell0}$; colored lines). Also, affinity-splitting between the monovalent cell model (red line) and the higher-order multivalent cell models (blue, green, pink and violet lines) becomes apparent in the high $k_x$ range, depending on whether model parameter conditions exhibiting positive or negative cooperativity~\cite{watabe2019, wofsy1992a}. A further analysis of the parameter conditions is required to investigate physical sources that give rise to the affinity-splitting.

%

\section{Conclusion}
Many GPCRs in the cell membrane randomly collide with each other, spontaneously taking the form of various oligomers such as dimers, trimers and tetramers. The role of receptor oligomerization (or aggregation) in GPCR signaling activations, however, has been elusive to date. In this article, we constructed a multiary complex model to investigate biophysical effects arising from various unobserved aggregated receptor states in GPCR signaling activations. Our results from model simulations revealed that receptor oligomerization functions to largely vary the overall affinity of ligand-receptor binding in a regime which cannot be ruled by cooperativity factor in the simplest ternary complex model. 

A further challenge for our work is to include the modification to the multiary complex models, such as the transitions between inactive and active states of the receptor observables. This was required in the past when the original ternary complex models were modified to extended and cubic ternary complex models~\cite{kenakin2017}. Such model modification leads to a more general modeling framework of GPCR signaling activations, and is of relevance more broadly beyond receptor aggregation presented here. This generalization raises questions of how aggregation processes of active-inactive state receptors are biophysically coupled with other signaling properties, e.g., the amplification and propagation of noisy signals~\cite{ueda2007, shibata2005}, and the physical limit and sensitivity to chemical concentration sensing in ligand-receptor binding~\cite{mora2019, kaizu2014, hu2010, bialek2005, berg1977}. Our work sheds light on these interesting questions from the perspective of theoretical biophysics, and suggests concrete modeling principles to explore general rules of receptor aggregation governing signaling activities and properties in various signal transduction systems.

\begin{acknowledgments}
We would like to thank Yasushi Okada, Jun Kozuka, Michio Hiroshima, Kozo Nishida, Wei Xiang Chew, Suguru Kato, Toru Niina, Koji Ochiai, Keiko Itano, Kotone Itaya, Takuya Miura and Kaoru Ikegami for their guidance and support throughout this research work. 
\end{acknowledgments}

%

\end{document}